\documentclass[twocolumn,superscriptaddress,longbibliography]{revtex4-1} 

\usepackage{amssymb,amsmath}
\usepackage{color}
\usepackage{graphicx}
\usepackage{float}
\usepackage[dvipsnames]{xcolor}
\usepackage[pdftex,breaklinks,colorlinks,citecolor=blue,linkcolor=blue,urlcolor=blue]{hyperref}

\begin{document}

\title{Multichannel interactions of two atoms in an optical tweezer}

\newcommand{\affa}{\affiliation{Department of Chemistry and Chemical Biology, Harvard University, Cambridge, Massachusetts, 02138, USA}}
\newcommand{\affb}{\affiliation{Department of Physics, Harvard University, Cambridge, Massachusetts, 02138, USA}}
\newcommand{\affc}{\affiliation{Harvard-MIT Center for Ultracold Atoms, Cambridge, Massachusetts, 02138, USA}}
\newcommand{\affd}{\affiliation{Department of Physics and Astronomy,  University of Toledo, Mailstop 111, Toledo, Ohio 43606, USA}}
\newcommand{\affe}{\affiliation{Department of Chemistry, Purdue University, West Lafayette, Indiana, 47907, USA}}
\newcommand{\afff}{\affiliation{Department of Physics and Astronomy, Purdue University, West Lafayette, Indiana, 47907, USA}}

\author{J.D. Hood} \email[]{hoodjd@purdue.edu} \affe \afff
\author{Y. Yu} \affb \affa \affc
\author{Y.-W. Lin}  \affa \affb \affc
\author{J.T. Zhang} \affb \affa \affc
\author{K. Wang} \affb \affa \affc
\author{L.R. Liu} \affb \affa \affc
\author{B. Gao} \affd
\author{K.-K. Ni} \email[]{ni@chemistry.harvard.edu} \affa \affb \affc

\date{\today}
\begin{abstract}
The multichannel Na-Cs interactions are characterized by a series of measurements using two atoms in an optical tweezer, along with a multichannel quantum defect theory (MQDT) with minimal input parameters.  The triplet and singlet scattering lengths are measured by performing Raman spectroscopy of the Na-Cs motional states and least-bound molecular state in the tweezer. The two-scale MQDT improves accuracy over the single-scale model by incorporating the $-C_8/r_8$ potential in addition to the $-C_6/r_6$ potential. Magnetic Feshbach resonances are observed for only two atoms at fields which agree to within $1\%$ of the MQDT predictions.    
Our tweezer-based approach combined with an effective theory of interaction provides a new methodology for futures studies of more complex interactions, such as atom-molecule and molecule-molecule, and where the traditional high-phase-space density bulk-gas techniques are technically challenging.
\end{abstract}

\maketitle

\section{Introduction}
Tuning interactions in ultracold gases of atoms and molecules  via Feshbach resonances or optical lattices has enabled studies of many rich quantum phenomena such as the BEC-BCS crossover~\cite{Regal2004}, superfluid-to-Mott insulator transitions~\cite{Greiner2002a}, and supersolidity~\cite{Bottcher2019, Tanzi2019, Chomaz2019}.
Feshbach resonances have also been utilized to associate loosely-bound  molecules, which has been an an important step for creating  ultracold  ro-vibrational ground state molecules~\cite{Ni2008, Molony2014, Takekoshi2014, Park2015, Guo2016}.  A key prerequisite for these experiments is an understanding of the underlying two-body and few-body interactions.  Although the origins of these interactions are complex molecular potentials, in the low-temperature regime effective theories can describe the interactions with no reliance on the short-range potentials.  For example, the single-scale multichannel quantum defect theory (MQDT) can provide an efficient description of atom-atom interactions in all spin channels and partial waves using only three parameters: the triplet and singlet scattering lengths, and the van der Waals $C_6$ coefficient~\cite{Gao2005a}.

Probing two- to few-body interactions in cold atoms has traditionally been performed by scattering experiments in bulk gases~\cite{Weiner1999} or by spectroscopy in optical lattices~\cite{Ospelkaus2006c,Danzl2010,Covey2016,Goban2018,Amato-Grill2019}, both of which rely on initially preparing high phase-space-density (PSD) gases. 
More recently, single atoms trapped in optical tweezers have become widely pursued as a versatile experimental platform for studying few and many-body physics through bottom-up scaling~\cite{Schlosser2001, Darquie2005, Miroshnychenko2006, Yavuz2006, Kaufman2015, Thompson2013, Norcia2018, Saskin2019, Covey2019}.   
An optical tweezer with only two atoms is a pristine environment for studying ultracold collisions ~\cite{Xu2015, Liu2018, Guan2019} or for producing two-particle entanglement~\cite{Kaufman2015, Sompet2019}.   Optical tweezers are now even being used with single ultracold molecule assembly \cite{Liu2018,Liu2019} and trapping~\cite{Anderegg2019}.

\begin{figure}[ht!]
\centering
\includegraphics[width=0.91\columnwidth]{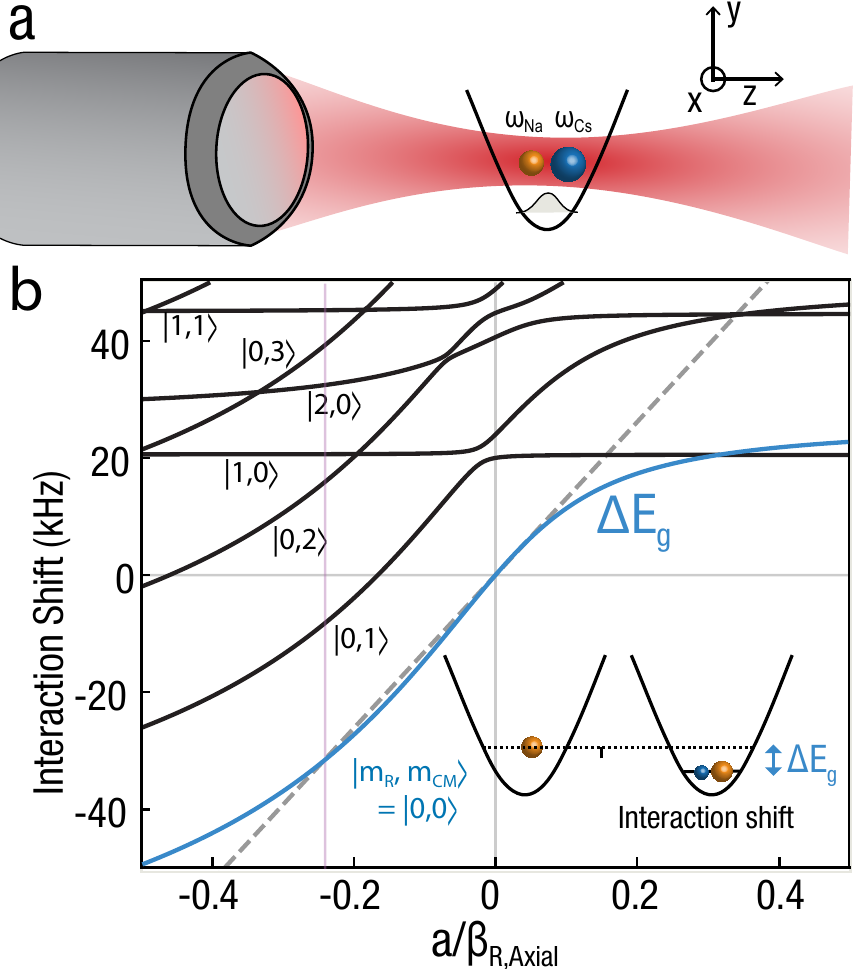}
\caption{\label{fig:calc}   
(a)  A single Na and Cs atom are trapped and cooled to the motional ground state of an optical tweezer.    (b)  The Na-Cs interaction shifts their motional trapping frequencies in the tweezer (see inset).  These trapping energies are calculated as a function of the scattering length , normalized by the relative oscillator length $\beta_{R,\text{Axial}} = 158$~nm, for various motional states, including the ground state $\Delta E_g$ (blue).  The ground state energy $\Delta E_g$ (blue) is shifted to higher (lower) frequencies for positive (negative) scattering lengths, corresponding to a repulsive (attractive) potential. The motional states with relative and center-of-mass axial excitations $|m_{R},m_{\text{CM}}\rangle$  are also plotted, while radial excitations are off the scale. 
}
\end{figure}

The natural questions arise: would measurements performed on two particles in an optical tweezer be sufficient to fully characterize two-body interaction, including the identification of Feshbach resonances? And, in the future, can such a platform offer a way to investigate interactions of more complex composite particles?

In this letter we probe the collisions of a single Na and a single Cs atom in an optical tweezer and in fully-controlled internal and external quantum states  without any contribution from multi-body effects or intra-species processes.  Despite previous characterization of the Na-Cs potential using Fourier-transform spectroscopy with hot atoms~ \cite{Docenko2006},  the near-threshold NaCs ground-state potential and Feshbach resonances have not been directly probed. Our work combines  Raman spectroscopy of trap motional states and the least-bound molecular state with a two-scale MQDT to extract the Na-Cs singlet and triplet scattering lengths.
The single-scale  MQDT~\cite{Gao2005a,Hanna2009,Gao2011,Makrides2014,Cui2018b} describes low-energy alkali interactions with the fewest parameters by separation the long-range potential $-C_6/r^6$ (with length scale $\beta_6 = (2 \mu C_6/\hbar^2)^{1/4}$) from the short-range potential, which is captured by the singlet and triplet scattering lengths. But its accuracy for magnetic Feshbach resonances decreases for systems with larger hyperfine splittings due to the larger energy scale~\cite{Makrides2014}.  We introduce a two-scale MQDT that captures the larger energy variation with a the shorter length scale  $-C_8/r^8$ potential. 
We observe the first Na-Cs Feshbach resonances, and the corresponding magnetic fields  agree to within  $1\%$ of our effective theory.

Our tweezer-based scheme can be extended to more complex interactions, such as atom-molecule or molecule-molecule interactions~\cite{Mayle2012,Mayle2013,Gregory2019}. The diffraction-limited optical tweezer creates a high effective density for collisions of around $\rho \approx 10^{14}\, \text{cm}^{-3}$, which is useful for species where achieving high densities is otherwise experimentally challenging. 
Whereas traditional bulk-gas and optical lattice experiments require suitable collisional properties (including miscibility) between all species in order to obtain a high PSD, the tweezer method instead achieves a high PSD by optically cooling individual particles~\cite{Thompson2013,Kaufman2015,Norcia2018,Covey2019} before merging them together.

\section{Calculation}
We perform our experiment with a single $^{23}$Na and a single $^{133}$Cs atom in the motional ground state of the same optical tweezer, as schematically shown in Fig.~\ref{fig:calc}(a) and reported in Ref.~\cite{Liu2019}. Initially, a single Na and Cs atom are  loaded stochastically into separate tweezers from a magneto-optical trap with a combined  probability of $\approx 35\%$~\cite{Liu2018}.  The final results, however, can be post-selected to guarantee both atoms are initially present with high confidence. The single atoms are then cooled simultaneously to their 3D motional ground state  using Raman sideband cooling~\cite{Monroe1995, Kaufman2012, Yu2018, Liu2019}. Subsequently, one of the atoms is transported and merged into the tweezer of the other atom, all while maintaining both atoms  in the  motional ground state~\cite{Liu2019, Wang2019}.  The resulting mean separation distance is 112 nm, which gives an effective density of $\rho = 2 \times 10^{14}\, \text{cm}^{-3}$. 

The strategy for extracting the triplet and singlet scattering lengths is outlined here.  The triplet least-bound binding energy is measured with two-photon spectroscopy and is directly related to the triplet scattering length by a single channel quantum defect theory (QDT)~\cite{Gao1998b,Gao2001,Gao2008a}, but extended to two-scale by including the $-C_8/r^8$ potential.  We also  measure the shifts of the Na-Cs motional states in the tweezer due to the interactions in various spin combinations, and relate the shifts to the scattering lengths through a numerical calculation of two atoms interacting in a harmonic potential.  Finally, we use the two-scale MQDT (which accounts for the hyperfine interaction) to extract the singlet scattering length consistent with the measured scattering lengths for the different hyperfine spins. 

When both atoms are in the same trap,  they interact  through their molecular potential $V(\mathbf{r}_1 - \mathbf{r}_2)$, which consists of  short-range molecular forces and a long-range van der Waals potential with a lowest-order term $-C_6/r^6$, where $r$ is the relative coordinate.  
In the low-energy limit where the de Broglie wavelengths are much larger than the molecular potential,  their interaction can be  well-modeled by a  Fermi pseudo-potential consisting of the scattering length $a$ and a regularized $\delta$-function,  $V(\mathbf{r}_1 - \mathbf{r}_2) = \frac{2\pi \hbar^2}{\mu} a \,\delta^{(3)}(\mathbf{r}_1 -\mathbf{r}_2)  \frac{\delta}{\delta r} r$  \cite{Busch1998}, where $\mu$ is the reduced mass.
The validity of the pseudo-potential is characterized by the ratio of the van der Waals length $\beta_6 = (2\mu C_6 / \hbar^2 )^{1/4}$ to the  relative harmonic oscillator lengths $\beta_R = \sqrt{\hbar/ \mu \omega_R }$~\cite{Bolda2002,Blume2002a}, where $\omega_R$ is the relative trapping frequency.  In our experiment, these are $\beta_6 \approx  6$ nm, $\beta_{R,\text{Radial}} \approx 66$ nm for the radial axes, and  $\beta_{R,\text{Axial}} \approx 158$ nm for the axial axis.

\begin{figure*}
\centering
\includegraphics[width=0.99\textwidth]{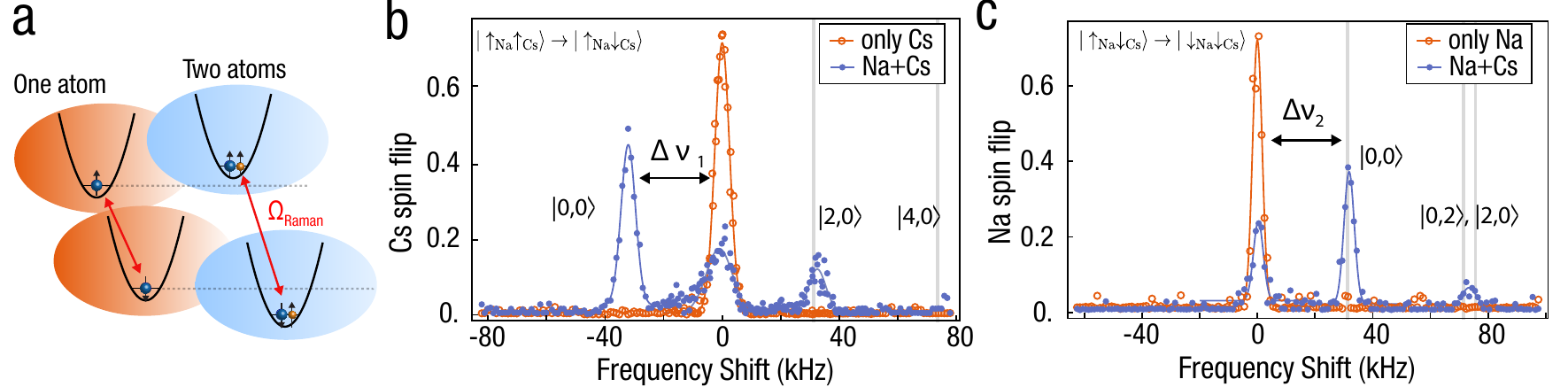}
\caption{\label{fig:int}  
(a) The  interaction shifts are measured with an optical Raman transition $\Omega_{\text{Raman}}$ that flips the hyperfine spin of one atom with (blue) and without (orange) the second atom present. The frequency difference contains the interaction shifts of both the initial and final state. 
\textbf{(b)} Raman spectrum for  ${|\!\uparrow_{\text{Na}}\uparrow_{\text{Cs}}\rangle}\rightarrow{|\!\uparrow_{\text{Na}}\downarrow_{\text{Cs}}\rangle}$ (blue solid),   as well as for flipping only the Cs hyperfine state ${|\!\uparrow_{\text{Cs}}\rangle}\rightarrow{|\!\downarrow_{\text{Cs}}\rangle}$ (orange open). 
The ${|0,0\rangle}$ peak is the shifted motional ground state, while the other peaks are excited motional states.  A peak remaining at zero frequency is due to population not initially in the ground state.  
(c)  Raman spectrum for  ${|\!\uparrow_{\text{Na}}\downarrow_{\text{Cs}}\rangle}\rightarrow{|\!\downarrow_{\text{Na}}\downarrow_{\text{Cs}}\rangle}$ (blue solid), as well as for the case of flipping only the Na hyperfine state ${|\!\uparrow_{\text{Na}}\rangle}\rightarrow{|\!\downarrow_{\text{Na}}\rangle}$  (orange open).
}
\end{figure*}

\begin{table}[t!]
\centering
\begin{tabular}{| c | c | c | }  
 \hline
 \hline
 Scattering length  &  This Work    &   Ref.~\cite{Docenko2006}     \\
 \hline
  $a_T$  &  30.4 $\pm$ 0.6  $a_0$ &   33 $\pm$ 5 $a_0$ \\
 $a_S$   &  428 $\pm$ 9 $a_0$ &   513 $\pm$ 250 $a_0$ \\
 \hline
 \hline
 Hyperfine Channel & Scattering length  & Interaction shift  \\
 \hline
Na(2,2)Cs(4,4),   ($\uparrow_{\text{Na}}\uparrow_{\text{Cs}}$)     & 30.4  $a_0$ &   1.40 kHz  \\ 
Na(2,2)Cs(3,3), ($\uparrow_{\text{Na}}\downarrow_{\text{Cs}}$)     &    -693.8 $a_0$   &  -30.7 kHz \\
Na(1,1)Cs(3,3), ($\downarrow_{\text{Na}} \downarrow_{\text{Cs}}$)  & 13.7 $a_0$   &  0.62 kHz \\
 \hline
 \hline
 Hyperfine Channel &  Feshbach resonance  &  MQDT Prediction \\
\hline
Na(1-1)Cs(3,-3) $s$-wave  &  652.1 $\pm$ 0.4 G  &  663 G   \\     
Na(1,-1)Cs(3,-3) $p$-wave &  791.10 $\pm$ .05 G   &  799 G   \\   
  \hline
\end{tabular}
    \caption{ Summary of measured scattering parameters.  
    The triplet $a_T$ and singlet $a_S$ scattering lengths are compared to the results in Ref.~\cite{Docenko2006}.   The measured scattering length for various hyperfine channels as well as the corresponding interaction shift.  The Feshbach resonances are compared to the two-scale MQDT predictions. 
}
\label{table:1}
\end{table}

The two-body interaction shifts  the motional trapping energies of the atoms in the tweezer, which is  calculated  as a function of the scattering length in Fig.~\ref{fig:calc}(b) and probed experimentally in Fig.~\ref{fig:int}.  
When the atoms have the same trapping frequencies, as in the case of identical atoms, the Hamiltonian is separable into center-of-mass~(CM) and relative coordinates, for which analytical results exist for a regularized $\delta$-function in a spherically~\cite{Busch1998} and cylindrically ~\cite{Idziaszek2006} symmetric harmonic trap. However, Cs trapping frequencies are $19 \%$ larger than Na trapping frequencies, which are $(\omega_{\text{Na},x},\,  \omega_{\text{Na},y} ,\, \omega_{\text{Na},z} ) = 2 \pi \times   (109, 118, 20)  $ kHz. We therefore calculate the shifted frequencies using the analytic solutions for the separable, cylindrically symmetric Hamiltonian, and then diagonalize a matrix containing the remaining non-separable ~\cite{Bertelsen2007,Deuretzbacher2008} and anisotropic terms  of the full Hamiltonian. The details are described in appendix~\ref{app:int}.

In Fig.~\ref{fig:calc}(b), the calculated Na-Cs motional trapping energies are plotted as a function of the scattering length $a$. The motional ground state $\Delta E_g$ (blue) shifts to higher (lower) frequencies for positive (negative) scattering lengths, corresponding to a  repulsive (attractive) potential. The dashed line is the first-order perturbation theory described in Appendix~\ref{app:pert}.
 For the excited motional states (black), states with an odd relative axial quantum number $m_{R}$ have no shift because the relative wavefunction is zero at the $\delta$-function, and therefore are non-interacting, while states with even $m_{R}$ are interacting.


\section{Interaction shift experiment} 
Experimentally, we measure the interaction shifts by resonantly flipping the ground hyperfine spin of one but not the other atom and then comparing it to the bare hyperfine splitting measured in the absence of the other atom, as shown in  Fig.~\ref{fig:int}(a).  
 Atoms are initially prepared in stretched state spin combinations that are stable against hyperfine changing spin collisions and  denoted by   ${|\!\uparrow_{\text{Cs}}\rangle}={|F\!=\!4,m_F\!=\!4\rangle_{\text{Cs}}}$,  
 ${|\!\downarrow_{\text{Cs}}\rangle}={|F\!=\!3,m_F\!=\!3\rangle_{\text{Cs}}}$, ${|\!\uparrow_{\text{Na}}\rangle}={|F\!=\!2,m_F\!=\!2\rangle_{\text{Na}}}$, and  ${|\!\downarrow_{\text{Na}}\rangle}={|F\!=\!1,m_F\!=\!1\rangle_{\text{Na}}}$.   
 The spin flip of one atom is driven by an optical Raman pulse with co-propagating beams at a bias  magnetic field of 8.8 gauss.

For the case of flipping a Cs spin in Fig.~\ref{fig:int}(b), several interaction-shifted peaks are observed by comparing ${|\uparrow_{\text{Na}}\uparrow_{\text{Cs}}\rangle}\rightarrow{|\uparrow_{\text{Na}}\downarrow_{\text{Cs}}\rangle}$  (blue)  to the one-atom case ${|\uparrow_{\text{Cs}}\rangle}\rightarrow{|\downarrow_{\text{Cs}}\rangle}$ (orange).
The largest peak is the shifted ground motional state ${|m_{R},m_{\text{CM}}\rangle}$=${|0,0\rangle}$.
The  shift gives the  difference of the interaction shifts of the initial and final spin configurations, $\Delta\nu_1 = ({\Delta E_g(\uparrow_{\text{Na}}\uparrow_{\text{Cs}})}-{\Delta E_g(\uparrow_{\text{Na}}\downarrow_{\text{Cs}})} )/\hbar =  -32$.1(2)kHz.  The smaller peaks at positive frequency shifts correspond to the motional excited states ${|2,0\rangle}$ and ${|4,0\rangle}$, which can be populated because they have some overlap with the initial state due to the wavefunction modification by the strong interactions.
The peak near zero frequency corresponds to the initial Na and Cs population that is not prepared in the motional ground state or an interacting state. The fitted height 0.46 of the $|0,0 \rangle$ peak serves as a lower bound for the relative motional ground state population.

Similarly, for the case of flipping a Na spin 
${|\uparrow_{\text{Na}}\downarrow_{\text{Cs}}\rangle}\rightarrow{|\downarrow_{\text{Na}}\downarrow_{\text{Cs}}\rangle}$ in Fig.~\ref{fig:int}(c), we observe several interaction shifted peaks corresponding to the motional states of ${|0,0 \rangle}$, ${|2,0 \rangle}$, and ${|0,2\rangle}$, and a non-shifted peak as the initial non-interacting population. 

\begin{figure}[t!]
\centering
\includegraphics[width=0.95\columnwidth]{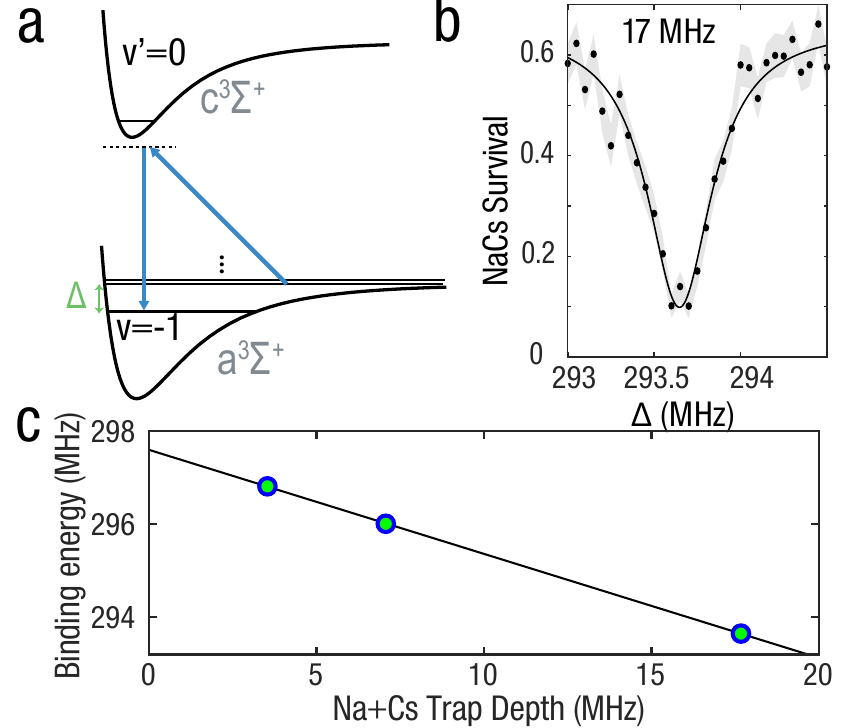}
\caption{\label{fig:binding} 
(a)  Raman spectroscopy to measure the binding energy of the least-bound state $v=-1$ of the ground molecular potential  $a^3\Sigma^+$. The tweezer light is also used for both branches of Raman light.
(b) Example Raman spectrum, where the two-photon relative detuning $\Delta$ is scanned until the molecular resonance is observed and both atoms are lost.
(c) Measured binding energies for triplet least-bound molecular state as a function of the tweezer power in units of Na+Cs trap depth. 
}
\end{figure}

Because interaction shifts only give the difference of the shifts between two states, we determine an absolute interaction shift of the triplet ${|\uparrow_{\text{Na}}\uparrow_{\text{Cs}}\rangle}$ by measuring the triplet least-bound ($v=-1$) binding energy, which can be related to the triplet scattering length directly through the two-scale single channel QDT. 
We measure the least-bound ($v=-1$) triplet binding energy with two-photon Raman spectroscopy, as schematically shown in Fig.~\ref{fig:binding}(a).   When the two-photon detuning is resonant with the binding energy, the atoms are transferred to the molecular state which is observed as simultaneous loss of both the Na and Cs atom~\cite{Liu2019}, as shown in the spectrum in Fig.~\ref{fig:binding}(b).  The resonance positions are plotted in~Fig.~\ref{fig:binding}(c) as a function of different tweezer powers in order to extrapolate the  binding energies without light shift.  For these experiments, the optical tweezer light is also used as the Raman beams, and is detuned $+18.2$ GHz from the $c^3\Sigma_{\Omega=1}(v\!=\!0,J\!=\!2)$ line at 288,698.2 GHz.   A linear extrapolation to zero power gives a $N=0$ binding energy of 297.6(1) MHz.   

The two-scale single-channel QDT relates the binding energies to the scattering lengths.  
For the van der Waals coefficients, we use the $C_6=3227$ a.u. and $C_8=3.681 \times 10^5$ a.u. from Refs.~\cite{Docenko2006, Derevianko2001, Porsev2003}.
The $N=0$  binding energy then gives a triplet scattering length $a_T = a(\uparrow_{\text{Na}}\uparrow_{\text{Cs}}) = 30.4(6) \, a_0$, where $a_0$ is the Bohr radius.  The scattering lengths are summarized in Table.~\ref{table:1}.

From the triplet scattering length from the binding energy measurement, the pseudo-potential model then gives an absolute interaction shift of $\Delta E_g(\uparrow_{\text{Na}}\uparrow_{\text{Cs}})/h$ = 1.40 kHz, which can be used to obtain an absolute shift for each state in the interaction shift measurements.    The measurement  $\Delta \nu_1$ in Fig.~\ref{fig:int}(b)  gives an absolute interaction shift  $\Delta E_g(\uparrow_{\text{Na}}\downarrow_{\text{Cs}}) /h = -30.7$ kHz, corresponding to a scattering length of $a(\uparrow_{\text{Na}}\downarrow_{\text{Cs}}) =  -693.8\, a_0$.   The two-scale MQDT is then used to   extract a singlet  scattering length of $a_{S} = 428(9)$ $a_0$. 

With our measured triplet $a_{T}$ and singlet $a_{S}$ scattering lengths, MQDT can now describe Na-Cs interactions in all spin configurations and all magnetic fields. As a verification, we compare the MQDT calculation to the Na spin-flip interaction shift $\Delta\nu_2=31.8(2)$~kHz  in Fig.~\ref{fig:int}(c). The shifts predominately come from ${E_g(\uparrow_{\text{Na}}\downarrow_{\text{Cs}}) /h} = -30.7$ kHz. The MQDT  predicts for the final state, ($\downarrow_{\text{Na}}\downarrow_{\text{Cs}}$), a scattering length of  $a(\downarrow_{\text{Na}} \uparrow_{\text{Cs}})$ = 13.7 $a_0$ and a corresponding interaction shift of 0.64 kHz, which would result in a calculated $\Delta\nu_2=31.34$~kHz, consistent with the measurement.

\section{Feshbach resonances} 
\begin{figure}
\centering
\includegraphics[width=0.99\columnwidth]{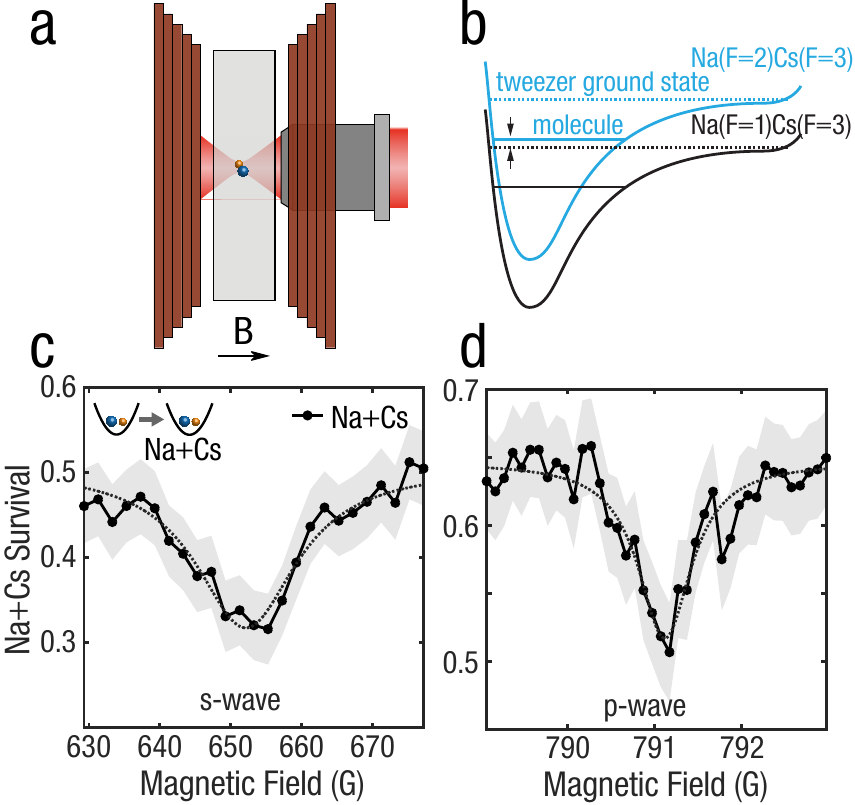}
\caption{\label{fig:feshbach}   (a) Apparatus for observing Feshbach resonances in an optical tweezer. The tweezer trap atoms in an ultra-high vacuum glass cell chamber between Helmholtz coils (orange), which produces a magnetic field $B$ of up to 1000 gauss at the atoms. (b)  Feshbach resonances occurs when the tweezer trap states  are tuned into resonance with a molecular bound state by applying a magnetic field. 
(c-d)  An $s$-wave and $p$-wave Feshbach resonance is observed by measuring the simultaneous loss of both Na and Cs as a function of magnetic field. The dashed line is a Gaussian fit.
}
\end{figure}

With Na-Cs interactions completely characterized at a low magnetic field, we use our two-scale MQDT to guide a search of  Feshbach resonances  at  much higher magnetic fields.
Feshbach resonances in ultracold atoms occur when pairs of atoms  are magnetically tuned into resonance with a closed molecular bound state, as shown in Fig.~\ref{fig:feshbach}(b). 
Searching for these resonances is typically done in bulk gases by measuring  atom loss near the resonances due to enhanced two-body or three-body inelastic collisions.  Here we demonstrate  such a search  starting with exactly two atoms in an optical tweezer.

Our search uses a separate optical tweezer apparatus, shown in Fig.~\ref{fig:feshbach}(a), with Helmholtz coils that are capable of producing a magnetic field $B$ up to 1000 G. The atoms are prepared in the lower hyperfine manifolds, Na($F\!=\!1$,$m_F\!=\!-1$) and Cs($F\!=\!3$, $m_F\!=\!-3$), at about 100 $\mu$K. Despite the high temperature as compared to typical bulk-gas-based searches, the tight confinement in the optical tweezer increases the sensitivity to collisional losses.  The B-field is ramped on, the atoms are merged into the same tweezer and held for 100 ms, and then the atoms are separated and imaged individually to check for loss. Because there are no lower energy states with the same total $M_F$,  the inelastic loss must occur through anisotropic interactions such as the electron spin-spin interactions.

Using our measured scattering parameters, the two-scale MQDT predicts two Feshbach resonances  for this hyperfine state, where the closed channel is either the rotational ground $N=0$ ($s$-wave) or excited $N=1$ ($p$-wave) of a molecular state which has the approximate quantum number $\nu=-1$ and Na($F\!=\!2$)Cs($F\!=\! 3$) in the low-field limit. Na-Cs loss spectroscopy is shown in Fig.~\ref{fig:feshbach}(c) and (d) for two different  magnetic field ranges.  The $s$-wave resonance is observed at 652.1(4) G, and the narrower $p$-wave Feshbach resonance is observed at 791.10(5) G,  which agree with the MQDT predictions with an accuracy of $1.4\%$ and $0.9\%$ respectively.
These are summarized in Table~\ref{table:1}.
With a colder sample for future study, inelastic confinement-induced resonances could shift the Feshbach resonance location due to coupling between center-of-mass and relative motion, as described in Ref.~\cite{Sala2012}.

\section{Conclusion}  
The agreement of the Feshbach resonance locations with the two-scale MQDT model combined with a series of tweezer-based measurements represents an important validation of the use of effective theory for interactions.   The interaction of composite particles, and in particular molecules, is generally too difficult for the standard coupled-channel approach because the number of coupled-channels is simply too large~\cite{Mayle2012,Mayle2013,Gregory2019}.  This difficulty requires that we look for effective theories to describe the interactions that require a minimum number of parameters, and efficient ways to measure these parameters. This tweezer scheme can also be used to probe atom-molecule and molecule-molecule interactions, providing both high effective densities and exact preparation of the collisional partners.
The measurement of the interaction shift of an atom in the presence of a molecule could also be used for non-destructive state-sensitive detection of the molecule, which is a challenge for molecules without closed optical transitions.  For Na-Cs in particular, the observation of a Na-Cs Feshbach resonance for exactly two atoms is an important step towards creating a single Feshbach molecule, and the eventual coherent creation of a ro-vibrational ground state molecule.

\appendix
\section{Two-scale multichannel quantum defect theory}
\label{app:mqdt}
The simplest description of the interaction between two alkali-metal atoms is a three-parameter model based on a single-scale multichannel quantum defect theory (MQDT) \cite{Gao2005a,Hanna2009,Gao2011,Makrides2014,Cui2018b}. Here ``single-scale'' corresponds to the fact the theory is built on the solutions for the long-range potential $-C_6/r^6$ which has a single length scale $\beta_6=(2\mu C_6/\hbar^2)^{1/4}$. And the three parameters can be taken as the singlet $s$-wave scattering length $a_S$, the triplet $s$-wave scattering length $a_T$, and the $C_6$ coefficient. The theory provides the simplest description of low-energy alkali interactions with or without magnetic field, including magnetic Feshbach resonances in all partial waves.

Depending on the specific system, the description of the magnetic Feshbach spectrum by this simple model has a typical accuracy of a few percent to about 8\% for heavier systems with large hyperfine splittings \cite{Cui2018b}. To understand these deviations and their improvements, one can keep in mind that an accurate description of a magnetic Feshbach resonance requires a \textit{simultaneous} accurate descriptions of both the open and the closed channels, in particular the bound spectrum in the closed channels over an energy range corresponding to the channel energy spacing. For alkali-metal atoms, the channels are separated by the hyperfine splittings, and the requirement for accurate Feshbach spectrum translates into the requirement of accurate bound spectrum over a binding energy range of $|\Delta E^{\mathrm{hf}}|$. Thus even when we are at ultracold temperatures where a single scattering length would suffice to describe the interaction in the open channel, the understanding of the scattering length itself, its relation to scattering lengths in other channels, and its tuning through the Feshbach spectrum, would require an underlying understanding over a much broader range of energies of at least $|\Delta E^{\mathrm{hf}}|$. The validity and the accuracy of single-scale MQDT depends on the validity that the energy variation over this energy range is due solely to the $-C_6/r^6$ potential, which is less valid for systems with large hyperfine splittings such as our system with a Cs atom that has a hyperfine splitting of around 9 GHz or 0.4 K.

There are different options for improving upon this single-scale baseline result. Focusing on effective theories that do not rely on the details of the short-range potential, the easiest and the simplest improvement to implement is to introduce parameters that characterize the energy dependence and the partial-wave dependence of the single-scale short-range parameters \cite{Li2014,Li2015}. 
A more fundamental approach that requires the fewest number of extra parameters, an important criteria for a good effective theory of interaction, is to go to a shorter length scale through the inclusion of higher order terms in the asymptotic expansion for the potential, such as
\[
-\frac{C_6}{r^6}-\frac{C_8}{r^8} \;,
\]
which is used as the reference potential for our 2-scale MQDT and is a more accurate representation of the real atomic interaction at shorter distances. The additional $-C_8/r^8$ terms has a corresponding length scale $\beta_8=(2\mu C_n/\hbar^2)^{1/(n-2)}=(2\mu C_8/\hbar^2)^{1/6}$ that is smaller than $\beta_6$. 

\begin{figure}
\includegraphics[width=0.99\columnwidth]{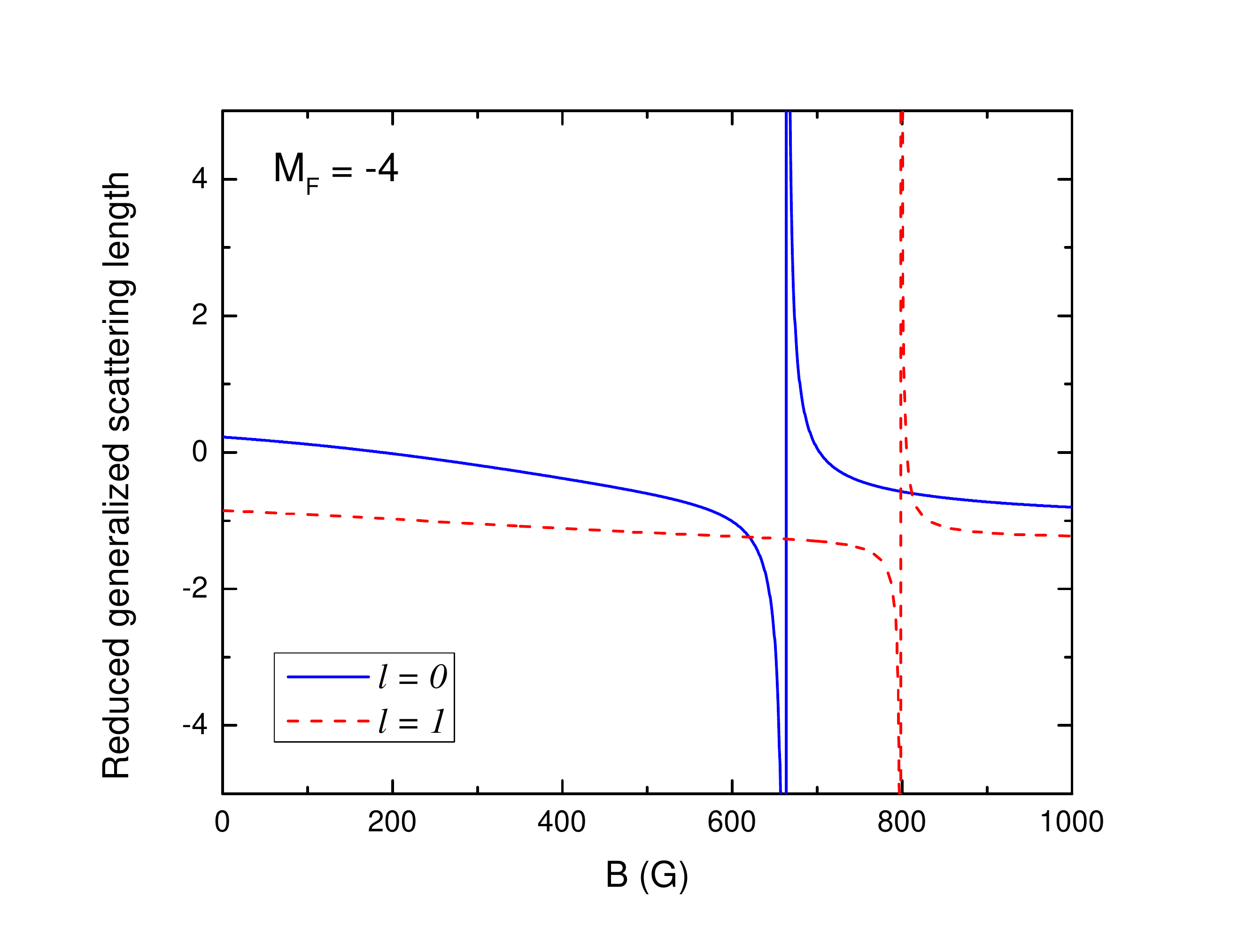}
\caption{The reduced generalized scattering lengths \cite{Gao2011,Makrides2014} for $s$ and $p$ partial waves in the Na$(-1)_1$Cs$(-3)_1$ channel. The $s$-wave has a Feshbach resonance at 663 G, and the $p$-wave has a Feshbach resonance at 799 G.
\label{fig:NaCsgslBMFm4}}
\end{figure}

A 2-scale MQDT description of alkali-metal interactions thus uses only one more parameter, the $C_8$, than the single scale theory for a total of 4 parameters that can be taken as $a_S$, $a_T$, $C_6$, and the $C_8$. It is more accurate in its description of scattering lengths and Feshbach spectrum, and can cover a greater range of energies when needed, similar to what has been demonstrated in QDT for $-C_1/r-C_4/r^4$ potential \cite{Fu2016a}. The theory is formally the same as the single-scale theory except for the details of the QDT functions. These details will be presented elsewhere. Figure~\ref{fig:NaCsgslBMFm4} shows the reduced generalized scattering lengths \cite{Gao2011,Makrides2014} for $s$ and $p$ partial waves in the Na$(-1)_1$Cs$(-3)_1$ channel over a $B$ field range of 0-1000 G, giving a more complete picture of the Feshbach resonances that we have observed. Table~\ref{tab:Na23Cs133FeshMFm4} gives the calculated properties of the Feshbach resonances.

We point out that similar theoretical analysis can also be carried out using numerical formulations of MQDT \cite{Mies1984a,Mies1984b,Burke1998,Ruzic2013} and other numerical methods of similar spirit which are already multiscale (see, e.g. \cite{vanKempen2002,Pires2014}). The main difference of our approach is that it tries to make the best use of the existing single-scale analytic solutions \cite{Gao1998a}, and is formulated to show explicitly how physics at different length scales are related, and the progression from the single-scale theory that is sufficient at small energies and requires fewer parameters, to multiscale theories at shorter length scales that requires more parameters. In such an approach, an extra parameter or the complexity of the theory is added only when necessary.  This characteristic is important in the context of an effective theory of interactions, especially when the parameters of the theory need to be determined from a limited number of experimental measurements.

For our particular system of $^{23}$Na-$^{133}$Cs, we have taken the long-range potential parameters $C_6 = 3227$ a.u. and $C_8 = 3.681\times 10^5$ a.u. from Docenko \textit{et al}. \cite{Docenko2006}. They are much more reliably determined by theory \cite{Derevianko2001,Porsev2003} than the parameters $a_S$ and $a_T$, which are sensitive to short-range complex molecular interactions. With still a limited number of experimental measurements, only the latter parameters are determined experimentally as discussed in the main text. With more future measurements, especially those related to more deeply bound molecular states, we anticipate more precise determinations of all parameters including the $C_6$ and $C_8$ coefficients in a future study.

\begin{table*}
\caption{Magnetic Feshbach resonances and their parameters for $^{23}$Na-$^{133}$Cs in the Na$(-1)_1$Cs$(-3)_1$ channel. } 
\begin{ruledtabular}
\begin{tabular}{c c c c c c c c c c c}
Channel& $M_F$ & $\ell$ & $B_{0\ell}$ (G) & $\Delta_{B\ell}$ (G) & $\widetilde{a}_{\text{bg}\ell}/\bar{a}_{\ell}$ & $\delta \mu_\ell / \mu_B$ & $K^{c0}_{\text{bg}\ell}$ & $g_{\text{res}}$ & $d_{B\ell}$ (G) & $\zeta_{\text{res}}$  \\ \hline 
$(-1)_1(-3)_1$ & $-4$ & 0 & 663.0 & 42.54 & -0.6487 & -0.6124 & -0.6065 & 2.366 & -16.74 & 1.300 \\ \hline 
$(-1)_1(-3)_1$ & $-4$ & 1 & 799.0 & 6.473 & -1.242 & -0.5026 & -4.128 & 3.850 & -33.19 & -0.1865 \\ \hline 
\end{tabular}
\end{ruledtabular}
\caption{A full explanation of the parameters can be found in Refs.~\cite{Gao2011, Makrides2014}.}
\label{tab:Na23Cs133FeshMFm4}
\end{table*}

\section{Interaction shift calculation}
\label{app:int}
 In the limit of small collision energy, the molecular potential $V(r)$ can be replaced with a  Fermi pseudopotential consisting of the scattering length $a$ and a regularized $\delta$-potential, 
\begin{equation}
V(|\mathbf{r}|) =   \frac{2 \pi \hbar^2 a}{\mu } \delta^{(3)}(\mathbf{r}) \frac{\partial}{ \partial r} r.
\end{equation}
For low energies, the Fermi pseudopotential reproduces the correct wavefunction outside the range of interactions.
The Fermi pseudopotential approximation accuracy is characterized by the ratio of the characteristic van der Waal's length $\beta_6 = (2\mu C_6 / \hbar^2 )^{1/4}$  to the relative oscillator length $\beta_R = \sqrt{\hbar/ m \omega_R}$~\cite{Bolda2002,Blume2002a}.   The Na mass is 23 amu and Cs mass is 133 amu. The measured trapping frequencies are  $(\omega_{\text{Na},x},\,  \omega_{\text{Na},y} ,\, \omega_{\text{Na},z} ) = 2 \pi \times   (109, 118, 20)  $ kHz,   and  $(\omega_{\text{Cs},x},\,  \omega_{\text{Cs},y} ,\, \omega_{\text{Cs},z} ) = 2 \pi \times   (130, 140, 24)$ kHz.   In our experiment,  $\beta_6 = 6$ nm is much smaller than the relative 
oscillator lengths $\beta_{R,\text{Axial}} = 158$ nm and $\beta_{R,\text{Radial}} = (65,67)$ nm.

The Hamiltonian for two different atoms interacting in an anisotropic 3D harmonic trap  is given by 
\begin{equation}
\begin{split}
H = \sum_{i=x,y,z}   \left( \frac{1}{2} m_1 \dot{x}^2_{1, i}  + \frac{1}{2} m_1 \omega^2_{1,i}  x_{1, i}^2 \right) \\
+ \sum_{i=x,y,z}  \left( \frac{1}{2} m_2 \dot{x}^2_{2, i}  + \frac{1}{2} m_2 \omega^2_{2,i}  x_{2, i}^2 \right)  
+  V(|\mathbf{r}_1 - \mathbf{r}_2|),
\end{split}
\end{equation}
where   $V(|\mathbf{r}_1 - \mathbf{r}_2|)$  is the spherically symmetric molecular potential for the two atoms.  The trapping frequencies are in general different for the two atoms due to different polarizabilities and masses. The trapping frequencies can also be different for all three axes in an anisotropic trap.     In terms of the relative and center-of-mass coordinates,   $\mathbf{r}_R = \mathbf{r}_1 - \mathbf{r}_2$  and $ \mathbf{r}_{C} = \frac{m_1 \mathbf{r}_1 + m_2 \mathbf{r}_2}{m_1 + m_2}$, the Hamiltonian can be re-expressed in terms of the reduced mass $\mu = \frac{m_1 m_2}{m_1 + m_2}$ ,  total mass  $M = m_1 + m_2 $,   relative frequencies  $ \omega_{R,i} =\sqrt{\frac{ m_2 \omega_{1,i}^2 - m_1 \omega_{2,i}^2}{m_1 + m_2}}$,  and center-of-mass frequencies $\omega_{C,i} = \sqrt{\frac{m_1 \omega_{1,i}^2 + m_2 \omega_{2,i}^2}{m_1 + m_2}}$  as  
\begin{equation} \label{eq:H1}
\begin{split}
H = \sum_{i=x,y,z}   \left( \frac{1}{2} M \dot{x}^2_{C, i}  + \frac{1}{2} M \omega^2_{C,i}  x_{C, i}^2 \right)\\
+ \left(  \sum_{i=x,y,z}   \left(\frac{1}{2} \mu \, \dot{x}^2_{R, i}  + \frac{1}{2} \mu \,\omega^2_{r,i}  x_{R, i}^2\right)  +V_{\text{int}}(|\mathbf{r}_R|)\right)  \\
+ \sum_{i=x,y,z}    \mu (\omega_{1,i}^2 - \omega_{2,i}^2) x_{R,i} x_{C, i}  .
\end{split}
\end{equation}
 The center-of-mass part of the Hamiltonian is just a single particle 3D harmonic oscillator.  The second term, the relative Hamiltonian,  is a 3D harmonic oscillator with a regularized $\delta$-function at the origin, for which  analytic solutions exist for cases that the trap has  spherical \cite{Busch1998}  and cylindrical \cite{Idziaszek2006} symmetries. 
 
The final term in the Hamiltonian in Eq.~\ref{eq:H1} couples  the center-of-mass and relative coordinates.  If the trapping frequencies of the two atoms are the same for each axis (for example in the case when the two atoms are the same species),  then the last term vanishes, and then the relative and center-of-mass coordinates are separable.   But when the two atoms are different species, as in our case, then this term can be important, which is discussed for the spherical case in Refs.~\cite{Bertelsen2007, Deuretzbacher2008}.
The trapping frequencies depend on the polarizability and mass as  $\sqrt{\alpha(\lambda )/ m}$.  At the tweezer wavelength of $\lambda$ = 976 nm, the measured Cs trapping frequencies are $\approx 19\%$ larger than those of Na.  

To find the eigenenergies of the full Hamiltonian of Eq.~\ref{eq:H1},  we first ignore the last term so that the Hamiltonian is separable into center-of-mass and relative coordinates.   In our experiment, the axial trapping frequency (along the axis of the tweezer and labeled as $z$) is approximately 5.6 times smaller than the two radial trapping frequencies ($x$ and $y$).  There is also a $7 \%$ difference between the two radial axes, but we initially assume they are the same and add the difference later as a correction.     Therefore, we use the cylindrical harmonic oscillator wavefunctions as the solutions of the center-of-mass Hamiltonian, and they are labeled as $|n,l,m_z\rangle$,  where $n$ and $l$ are the principal and angular momentum quantum numbers for the radial part, and $m_z$ is  quantum number for 1D harmonic oscillator for the axial part, and have eigenenergies
\begin{equation}  \label{eq:Ecylin}
E_{n,l,m_z}/ (\hbar \omega_z)=  ( 2 n + |l| + 1) \eta   +   (m_z + 1/2),   
\end{equation}
where we define $\eta$ as the ratio of the radial to axial trapping frequency.

For the relative Hamiltonian, we use the analytic cylindrical solutions  from Ref.~\cite{Idziaszek2006}.    These solutions require that the axial trapping frequency is an integer multiple $\eta$ of the radial trapping frequency.  We define $\eta = 6$, which is close to the actual values of $5.6$, and we will include  the remaining terms later as a correction.    The analytic solutions are given for the interacting states, but there also many relative states which have zero wavefunction at the $\delta$-function, and therefore are unaffected.  For example any state with $l\ne0$ or  odd $m_z$ has a zero at $\delta$-function.   The complete basis includes both the interacting states from Ref.~\cite{Idziaszek2006}  as well as all of the non-interacting states.  The non-interacting states are solutions to the cylindrical harmonic oscillator, and so are just cylindrical harmonic oscillator wavefunctions.   

One complication is that when $\eta$, the ratio of the radial to the axial trapping frequency, is an integer,  there is a subspace of cylindrical harmonic oscillator states  with $l=0$ and even $m_z$ that are degenerate and have the same energy from Eq.~\ref{eq:Ecylin}.   In each degenerate subspace with $N_{\text{deg}}$ states,  the non-interacting states are a linear superposition of the degenerate eigenstates $\psi_i$.  We  find these amplitudes $c_i$ using a Gram-Schmidt procedure, which requires that $\sum_{i=1}^{N_{\text{deg}}}  c_i \psi_i(0)$ = 0.In each subspace, there is only one interacting state, for which the analytic solution is used, and  $N_{\text{deg}}-1$ non-interacting states. 

For the interacting states, the energies are given by the transcendental equations~\cite{Idziaszek2006}
\begin{equation}
\mathcal{F}(-(E-E_0)/2 , \eta)  =   -\sqrt{2\pi}/a,
\end{equation}
where $\mathcal{F}(x,\eta)$ is given by 
\begin{equation}
\begin{split}
\mathcal{F}(x, \eta) =      \frac{\sqrt{\pi} \Gamma(x)}{\Gamma(x+\frac{1}{2})}  \sum_{m=1}^{n-1} F(1,x;x+\frac{1}{2} ; e^{i(2\pi m/ \eta )} ) \\ -      \frac{2 \sqrt{\pi} \Gamma(x)}{\Gamma(x-\frac{1}{2})} .
\end{split}
\end{equation}
Here $F(a,b;c,x)$ denotes the hypergeometric function and $\Gamma(x)$ is the Euler gamma function.  The energy $E$ and $E_0$ are in units of the axial trap energy $\hbar \omega_z$, and so the ground state energy $E_0  = \eta + 1/2$.

Now that we have solution in the separable and cylindrical case, the next step is to include the non-separable and asymmetric correction terms  by diagonalizing the total matrix in the combined center-of-mass and relative cylindrical bases.   For the matrix, we include all states with energies up to  20 $\omega_{R,z}$.    The matrix elements are calculated numerically using the cylindrical wavefunctions, which for completeness are given here:

\begin{equation}
\Psi_{n,l,m_z} (\rho,\theta,z) =  \Psi^{\text{radial}}_{n,l}(\rho, \theta)    \Psi^{\text{axial}}_{m_z}(z) ,
\end{equation}
with the normalized radial harmonic oscillator wavefunction
\begin{equation}
\begin{split}
  \Psi^{\text{radial}}_{n,l}(\rho, \theta) =   \sqrt{\frac{2  n!}{a_\perp^2 (n+|l|)!}}  e^{-r^2/(2 a_\perp^2)} ( r/ a_\perp)^{|l|} \\ \times  L_n^{|l|}( r^2/ a_\perp^2)    \frac{ e^{i l \theta}}{\sqrt{2 \pi}},
  \end{split}
\end{equation}
and the normalized 1D harmonic wavefunction 
\begin{equation}
  \Psi^{\text{axial}}_{m_z}(z) = \frac{1}{\sqrt{2^{m_z} m_z!}}  \frac{1}{\sqrt{a_z}(\pi)^{1/4}}   e^{-  z^2/(2 a_z)} H_{m_z}( z/a_z).
\end{equation}
Here the radial and relative oscillator lengths are defined as $a_\perp = \sqrt{\hbar/( \mu \omega_\perp)}$ and $a_z = \sqrt{\hbar/ (\mu \omega_z)}$.   $H_{m_z}$ are the Hermite-Gaussian functions, and $L^{|l|}_n$ are the generalized Laguerre polynomials. 

The eigenenergies of the matrix are calculated as a function of the scattering length, which is shown in Fig.~1 of the main text.

\section{Perturbation theory for ground state shift}
\label{app:pert}
We use first-order perturbation theory to estimate the shift of the Na-Cs 3D motional ground state due to the interaction, which is approximated by the  Fermi pseudopotential interaction $V(| \mathbf{r}_1 - \mathbf{r}_2|) =   \frac{2 \pi \hbar^2 a}{\mu } \delta^{(3)}(\mathbf{r}_1 - \mathbf{r}_2)  \frac{\partial}{ \partial r} r$.     Using the  Cartesian harmonic oscillator bases $|n^{\text{Na}}_x, n^{\text{Na}}_y, n^{\text{Na}}_z ; n^{\text{Cs}}_x, n^{\text{Cs}}_y, n^{\text{Cs}}_z \rangle $, first-order perturbation theory gives a ground state shift of $\Delta E_g \approx  \frac{2\pi\hbar^2 a}{\mu }\langle 0,0,0;0,0,0|\delta^{(3)}(\mathbf{r}_1-\mathbf{r}_2) \frac{\partial}{ \partial r} r  |0,0,0;0,0,0\rangle  $, which simplifies to  
\begin{equation} \label{eq:pert2}
\Delta E_g \approx   a   \left( \frac{2 \hbar^2}{\mu \sqrt{\pi}} \frac{1}{ \beta_{\text{eff}}^3  } \right) ,
\end{equation}
where the effective 3D oscillator length is defined in terms of the 1D oscillator lengths, 
\begin{equation}
\beta_{\text{eff}} =  \left( (\beta_{\text{Na},x}^2 +\beta_{\text{Cs},x}^2)(\beta_{\text{Na},y}^2 +\beta_{\text{Cs},,y}^2)(\beta_{\text{Na},z}^2 +\beta_{\text{Cs},,z}^2) \right)^{1/6}.
\end{equation}
The oscillator lengths are  $\beta = \sqrt{\hbar/ m \omega} $, where $m$ is the mass of the atom and $\omega$ is the trapping frequency.

\begin{acknowledgments}
We thank Lewis Picard, Eliot Fenton, and Frederic Condin for experimental assistance.
This work is supported by the Arnold and Mabel Beckman Foundation, as well as the NSF (PHYS-1806595 and through Harvard-MIT CUA), AFOSR (FA9550-19-1-0089), and the Camille and Henry Dreyfus Foundation. J. T. Z acknowledges support from an NDSEG fellowship.
The work at Toledo was supported by NSF (PHY-1607256).
\end{acknowledgments}


%

\end{document}